\documentclass[preprint]{elsarticle}

\usepackage{lineno}
\modulolinenumbers[5]

\usepackage{graphicx} 
\usepackage{epstopdf}
\usepackage{amssymb}
\usepackage{amsmath} 
\usepackage{float} 
\usepackage{natbib}
\usepackage[normalem]{ulem}

\usepackage[ruled,vlined]{algorithm2e}
\SetKwProg{Timestep}{Time Step}{}{end Time Step}
\SetKwProg{Momentfitting}{Moment Fitting}{}{end Moment Fitting}
\SetKwProg{Boundarymomentfitting}{Boundary Moment Fitting}{}{end Boundary Moment Fitting}

\DeclareMathOperator{\sech}{sech}

\newcommand{\D}{\mathrm{d}}

\usepackage[hidelinks]{hyperref}

\usepackage{xcolor}

\newcommand{\ib}{\ensuremath{\mathbf{i}}}
\newcommand{\xb}{\ensuremath{\mathbf{x}}}
\newcommand{\vb}{\ensuremath{\mathbf{v}}}

\makeatletter
\def\ps@pprintTitle{%
  \let\@oddhead\@empty
  \let\@evenhead\@empty
  \let\@oddfoot\@empty
  \let\@evenfoot\@oddfoot
}

\makeatother
\journal{---}

\begin{document}
\sloppy

\begin{frontmatter}

\title{A Parallel Low-Rank Solver for the Six-Dimensional Vlasov-Maxwell Equations}

\author[rub]{F. Allmann-Rahn\corref{correspondingauthor}}
\cortext[correspondingauthor]{Corresponding author}
\ead{far@tp1.rub.de}

\author[rub]{R. Grauer}

\author[uu]{K. Kormann}

\address[rub]{Ruhr University Bochum, Universitätsstraße 150, 44801 Bochum, Germany}
\address[uu]{Uppsala University, P.O.~Box 256, SE-751 05 Uppsala, Sweden}

\begin{abstract}

Continuum Vlasov simulations can be utilized for highly accurate modelling
of fully kinetic plasmas. Great progress has been made recently
regarding the applicability of the method in realistic plasma configurations.
However, a reduction of the high computational cost that is inherent
to fully kinetic simulations would be desirable, especially at high velocity space
resolutions. For this purpose, low-rank approximations can
be employed. The so far available low-rank solvers are restricted to
either electrostatic systems or low dimensionality and can therefore
not be applied to most space, astrophysical and fusion plasmas.
In this paper we present a new parallel low-rank solver for the full six-dimensional
electromagnetic Vlasov-Maxwell equations with a compression
of the particle distribution function in velocity space.
Special attention is paid to mass conservation and Gauss's law.
The low-rank Vlasov solver is applied to standard
benchmark problems of plasma turbulence and magnetic reconnection
and compared to the full grid method. It
yields accurate results at significantly reduced computational cost.\\

\end{abstract}


\end{frontmatter}


\section{Introduction}
\label{sec:introduction}

Fully kinetic continuum Vlasov simulations feature a highly accurate representation
of velocity space and are utilized in the modelling of plasmas with great success
(e.g.\ \cite{schmitz-grauer:2006-1,allmann-rahn-lautenbach-grauer:2021,pezzi-liang-juno-etal:2021,
juno-hakim-tenbarge-etal:2018,pezzi-cozzani-califano-etal:2019,pusztai-juno-brandenburg:2020}).

One remaining issue is that the possible resolutions are still very limited, since computations involve
the six-dimensional plasma distribution function. In order to reduce computational cost and reach higher resolutions,
low-rank techniques have been introduced to plasma physics in Kormann (2015) \cite{kormann:2015}
where the discretized plasma distribution function is approximated by a tensor of low rank.
This approximation can be considered a compression of the distribution function with
both lossy and lossless elements. The distribution function is decomposed by means 
of singular value decompositions (SVDs) and the information associated with small singular
values is discarded. The lossless aspect of the compression results from the exploitation
of separability.

In general, low-rank techniques have become very popular in the last years in many
areas of physics, mathematics, computer science and more
(see e.g.~\cite{schollwoeck:2011,cirac-etal:2021,bachmayr-schneider-uschmajew:2016,grasedyck-kressner-tobler:2013} and references therein).
In plasma physics and hydrodynamics different approaches to adapting tensor compression have been developed.
The approach from Ehrlacher \& Lombardi (2017) \cite{ehrlacher-lombardi:2017} starts
from the Hamiltonian formulation of the Vlasov-Poisson
system to find a natural splitting for the low-rank approximation. They use the proper generalized decomposition (PGD)
where the alternating least squares (ALS) optimization method is employed to find the low-rank tensor.
This is in contrast to SVD-based low-rank approximations where no optimization
problem is solved. In both \cite{kormann:2015} and \cite{ehrlacher-lombardi:2017} the high-dimensional computations from usual
Vlasov schemes are replaced by computations in low-rank format. During
each mathematical operation, the rank increases and is truncated back to lower rank.
A different method is introduced in Einkemmer \& Lubich (2018) \cite{einkemmer-lubich:2018} for the
Vlasov-Poisson system. They apply a dynamical low-rank approximation where 
the Vlasov equation is projected to a low-rank manifold so that all
computations are done within this manifold and the rank does not increase by the computations.
Position space and velocity space are decomposed so that the six-dimensional problem is
reduced to two three-dimensional advection problems times the rank. Alternatively, the
six-dimensional advection is split up into six one-dimensional advection problems times
the rank. The method is extended in \cite{einkemmer-lubich:2019} to conserve
mass, momentum and energy making use of a Lagrange-multiplier technique to enforce the
conservation properties. An extension from the electrostatic system to the 1D2V Vlasov-Maxwell
system is given in \cite{einkemmer-ostermann-piazzola:2020}. The Vlasov-Poisson
version of the code is ported to graphics processing units (GPUs) in \cite{cassini-einkemmer:2021}.
As in plasma physics, low-rank algorithms can also be utilized in the related field
of hydrodynamics.  Dolgov and Stoll (2017) \cite{dolgov-stoll:2017} apply low-rank approximations
to the optimization problems in the solution of Navier-Stokes equations.
Fluid dynamics can also utilize the Boltzmann equation with a
collision term which is done by means of low-rank decompositions in Einkemmer (2019) \cite{einkemmer:2019}.

The aforementioned low-rank Vlasov solvers are restricted to the electrostatic
Vlasov-Poisson system or in one case the electromagnetic Vlasov-Maxwell system in
one spatial dimension and two velocity dimensions. Therefore, they cannot yet be applied
to many of the physical problems arising for example in space and astrophysics or
in fusion research. In this paper we present a new parallel low-rank solver for the
full six-dimensional electromagnetic Vlasov-Maxwell system. We build upon the earlier
work in \cite{kormann:2015}, but choose to compress only the velocity space using the
hierarchical Tucker decomposition \cite{hackbusch-kuehn:2009,grasedyck:2010}. There are
multiple advantages with this approach as will be discussed in the next section, most
prominently this enables us to parallelize our algorithm for use on large high performance computing systems.

\section{Vlasov-Maxwell Equations and Low-Rank Decomposition}
\label{sec:vlasov_maxwell_low_rank}

A plasma is accurately characterized by six-dimensional
phase-space distribution functions. For numerical modelling we consider
the positional coordinate $\mathbf{x} \in \Omega_x \subset \mathbb{R}^3$
and the velocity coordinate $\mathbf{v} \in \Omega_v \subset \mathbb{R}^3$
so that $\Omega_x \times \Omega_v$ is a subset of the phase space,
and a time $t \in [0,\tau_{\text{max}}]$. Then the distribution function is
$f_{s}(\mathbf{x}, \mathbf{v}, t): \Omega_x \times \Omega_v \times [0,\tau_{\text{max}}] \rightarrow \mathbb{R}$
for each particle species $s$.
The particle motion in the collisionless case is governed by the Vlasov equation
\begin{equation}
\frac{\partial f_s(\mathbf{x}, \mathbf{v}, t)}{\partial t}
+ \mathbf{v} \cdot \nabla f_s(\mathbf{x}, \mathbf{v}, t)
+ \frac{q_s}{m_s} (\mathbf{E}(\mathbf{x}, t) + \mathbf{v} \times \mathbf{B}(\mathbf{x}, t))
\cdot \nabla_v f_s(\mathbf{x}, \mathbf{v}, t) = 0
\label{eq:vlasov}\end{equation}
with the electric field $\mathbf{E}$ and the magnetic field $\mathbf{B}$.
Together with Maxwell's equations
\[\nabla \cdot \mathbf{E}(\mathbf{x}, t) = \frac{\rho(\mathbf{x}, t)}{\epsilon_{0}}, \quad
\nabla \cdot \mathbf{B}(\mathbf{x}, t) = 0, \]
\begin{equation}\nabla \times \mathbf{E}(\mathbf{x}, t) = - \frac{\partial \mathbf{B}(\mathbf{x}, t)}{\partial t}, \quad
\nabla \times \mathbf{B}(\mathbf{x}, t) = \mu_{0} \mathbf{j}(\mathbf{x}, t) +
\mu_{0} \epsilon_{0} \frac{\partial \mathbf{E}(\mathbf{x}, t)}{\partial t}\end{equation}
the Vlasov equation provides a complete plasma description. The charge and current
densities are denoted by $\rho$ and $\mathbf{j}$, and $\mu_0, \epsilon_0$
are the vacuum permeability and the vacuum permittivity, respectively.

Plasma quantities in position space can be derived from the distribution function by taking moments,
i.e.\ multiplying $f_s$ by powers of $\mathbf{v}$ and taking the integral over velocity space.
The zeroth moment, the particle density is
${n_s(\mathbf{x}, t) = \int f_s(\mathbf{x}, \mathbf{v}, t) \text{d}\mathbf{v}}$ and the mean velocity (derived from
the first moment) is
${\mathbf{u}_s(\mathbf{x},t) = \frac{1}{n_s(\mathbf{x}, t)} \int \mathbf{v} f_s(\mathbf{x}, \mathbf{v}, t) \text{d}\mathbf{v}}$.
The second moment (multiplied by mass) is given by
${\mathcal{P}_{s}(\mathbf{x},t) =  m_{s} \int \mathbf{v} \otimes \mathbf{v} f_{s}(\mathbf{x}, \mathbf{v}, t) \text{d}\mathbf{v}}$,
where $\otimes$ denotes the Kronecker product. The charge and current densities can be obtained from
the moments as $\rho(\mathbf{x},t) = \sum_s q_s n_s(\mathbf{x},t)$ and
$\mathbf{j}(\mathbf{x},t) = \sum_s q_s n_s(\mathbf{x},t) \mathbf{u}_s(\mathbf{x},t)$.
Temperature ${\mathrm{T}}$ is related to the raw second moment ${\mathcal{P}}$ like
${\mathrm{T} = (\mathcal{P} - m n\,\mathbf{u}\otimes\mathbf{u})/(n k_B)}$.
Taking the zeroth moment and the first moment of the complete Vlasov equation leads to two fluid
equations, the continuity equation for particle density
\begin{equation}
\frac{\partial n_{s}(\mathbf{x},t)}{\partial t} + \nabla \cdot (n_{s}(\mathbf{x},t) \mathbf{u}_{s}(\mathbf{x},t)) = 0 \, ,
\label{eq:tenmoment_continuity} \end{equation}
and the momentum equation
\begin{equation}
m_{s} \frac{\partial (n_{s}(\mathbf{x},t) \mathbf{u}_{s}(\mathbf{x},t)) }{\partial t} =
n_{s} q_{s} (\mathbf{E}(\mathbf{x},t) + \mathbf{u}_{s}(\mathbf{x},t) \times \mathbf{B}(\mathbf{x},t)) -
\nabla \cdot \mathcal{P}_{s}(\mathbf{x},t).
\label{eq:tenmoment_movement} \end{equation}

Higher moments of the Vlasov equation yield further equations and this hierarchy forms the full set of fluid equations.
The hierarchy is exact but any finite truncation is not closed because the equation for one moment always includes
the next higher moment. Considering only the first two fluid equations, one is missing an equation for ${\mathcal{P}_{s}}$.
In Sec.~\ref{sec:dual_solver} we will obtain ${\mathcal{P}_{s}}$
from the distribution function in order to close the system and get exact kinetic results for $n_s$ and $\mathbf{u}_s$.

In this paper we present a solver for the Vlasov equation \eqref{eq:vlasov} that
makes use of low-rank approximations. There, the
central operation is the singular value decomposition (SVD), since a truncated singular value
decomposition based on the $r$ largest singular values corresponds to the best rank-$r$
approximation in the Frobenius norm and other norms that are invariant under unitary
transformations \cite{schmidt:1907,eckart-young:1936,mirsky:1960}.
If the singular values decay fast, the SVD provides a very efficient
approximation in the sense that a sufficiently exact reconstruction can be made although
only the most important information (contained in the large singular values) was kept. This can be
thought of as a lossy compression of the matrix.

There is also a loss-free aspect of this matrix compression
as the SVD is an efficient representation of separable matrices.
In the low-rank Vlasov solver this property
is exploited, for example distribution functions that are close to
Maxwellian can be represented with small ranks.
A plasma is typically Maxwell distributed in calm or collision-dominated regions.
The Maxwell distribution separates in velocity space like
$f_{\text{Maxwell}}(\mathbf{x}, \mathbf{v}) = f_{1}(\mathbf{x}, v_{x})\ f_{2}(\mathbf{x}, v_{y})\ f_{3}(\mathbf{x}, v_{z})$,
where $f_i$, $i \in \{1,2,3\}$, is a normal distribution with an expectation value of $\mu_i(\mathbf{x}) = u_i(\mathbf{x})$ and a standard deviation
of $\sigma (\mathbf{x}) = \sqrt{\frac{k_{B} T_{s}(\mathbf{x}) }{m_{s}}}$. The separable Maxwell distribution is of rank one in velocity space and
can therefore be compressed very well.
If three-dimensional velocity space $\Omega_v$ at position $\mathbf{x}$ is represented by a grid with $C$ cells in each
velocity direction, a Maxwellian distribution is fully represented by $3 C$ instead of $C^3$ values.
This is a strong motivation for using low-rank approximations to speed up computations that involve
the plasma distribution function.

\section{Low-Rank Vlasov Solver}

\subsection{Hierarchical Tucker Decomposition of the Distribution Function}
\label{sec:htd}

\begin{figure}
\centering \includegraphics[width=0.65\textwidth]{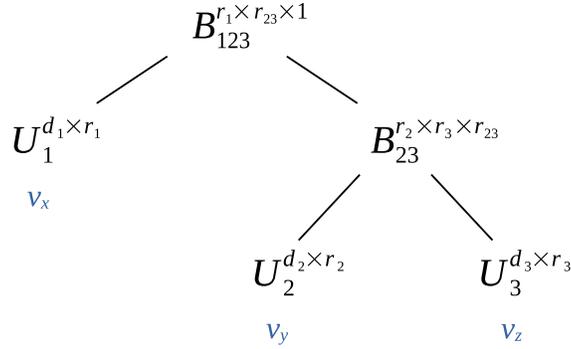}
\caption{At each coordinate $(x,y,z)$ the velocity distribution $f(v_x,v_y,v_z)$ is decomposed
  according to the shown hierarchical Tucker tree.
  The coordinates which the respective factor matrices belong to are highlighted in blue.}
\label{fig:ht_tree_f}\end{figure}

Numerical Vlasov solvers discretize the particle distribution function on
a six-dimensional phase space grid so that each cell in the grid corresponds
to a value of $f$ at a phase space point $(x,y,z,v_x,v_y,v_z)$ (or
the average value of $f$ in the cell around that point). The discretized $f$ can
be considered a sixth order tensor and low-rank tensor decomposition techniques
can be applied to $f$. In the previous section the low-rank decomposition of a
matrix was discussed. The concept can be generalized to tensors in various ways,
for example in the form of the higher-order singular value decomposition/Tucker
decomposition, the tensor train decomposition/matrix product states, or the
hierarchical Tucker decomposition (HTD).
Here, the hierarchical Tucker decomposition is chosen for the
low-rank approximation of $f$. It is computed by constructing
matricizations (also called flattenings) of the tensor that is supposed to be
decomposed. A matricization of the distribution function
$\mathrm{f} \in \mathbb{R}^{N_x \times N_{v_x} \times N_y \times N_{v_y} \times N_z \times N_{v_z}}$ can
for example be the matrix $M \in \mathbb{R}^{N_x N_{v_x} \times N_y N_{v_y} N_z N_{v_z}}$. Naturally,
it is possible to choose different ways to combine the dimensions in order to obtain a matrix.
This choice determines the tree structure of the HTD. To a matricization and its transpose the singular value
decomposition can be applied. Matrices that are part of the decomposition are then rearranged to
different matricizations and again SVDs are applied. This procedure is repeated hierarchically.
The computation of the HTD and mathematical operations
in HT format are described in the original papers by Hackbusch \& Kühn \cite{hackbusch-kuehn:2009} and
Grasedyck \cite{grasedyck:2010}. Detailed information on
the implementation in computer programs is given by Tobler and Kressner
\cite{kressner-tobler:2014,tobler-kressner:2012}.

When the low-rank method was first used for the electrostatic
Vlasov equation in \cite{kormann:2015}, the choice was to apply a tensor train decomposition
to the six-dimensional distribution function $f(x,y,z,v_x,v_y,v_z)$
to compress the full phase space information. This approach can be extended to
the Vlasov-Maxwell equations. However, it is so far not possible
to efficiently parallelize a Vlasov solver where all six dimensions of the
distribution function are compressed. Combined with the fact that high ranks are necessary
to represent large-scale physical configurations, the applicability of such a solver
for realistic physical problems is very limited.
There are several reasons why it is favourable to compress
only the velocity space at each point $(x,y,z)$ and
keep the grid in position space. Due to the separability of a Maxwellian distribution,
velocity space is expected to compress well and the worst case of rank one is
still physically sensible. There is more freedom in the
choice of advection schemes which is important for numerical stability.
In the case of compression in all six dimensions, element-wise products
in hierarchical Tucker format would need to be performed, whereas
computationally cheaper additions are sufficient when only velocity
space is compressed. Finally, efficient parallelization by means of a domain decomposition
approach in position space is straightforward.

The hierarchical Tucker tree of the distribution function is shown in Fig.~\ref{fig:ht_tree_f} in the form that we use.
It consists of factor matrices $\mathrm{U}_i \in \mathbb{R}^{N_i \times r_i}$
where $N_i$ is the number of velocity space cells and $r_i$ is the rank in each direction
$i \in \{1,2,3\}$. They are connected via the transfer tensors
$\mathrm{B}_{23} \in \mathbb{R}^{r_2 \times r_3 \times r_{23}}$
and $\mathrm{B}_{123} \in \mathbb{R}^{r_1 \times r_{23} \times 1}$.
For each point of the grid in position space, indexed by $\ib=(i_1,i_2,i_3)$, we have a hierarchical Tucker tensor of the form
\begin{equation}\label{eq:representationf}
	\text{vec}\left(f(x_{i_1}, y_{i_2},z_{i_3},:,:,:)
	\right) = \left( (U_3^{\ib} \otimes U_2^{\ib}) B^{\ib}_{23} \otimes U_1^{\ib}  \right) B^{\ib}_{123}.
\end{equation}

The original motivation for choosing the hierarchical Tucker decomposition
over the HOSVD/Tucker decomposition or tensor trains was that in the
compression of the full six-dimensional phase space, the ordering of
dimensions in the tensor train format is unintuitive \cite{kormann:2015}
whereas the HOSVD has a high-dimensional core tensor. In the new approach of
compressing only velocity space, the HTD can be exchanged with the
HOSVD or tensor trains with no major differences in the level of compression
to be expected. One advantage of the HTD may be that many plasma configurations
feature different physics in one direction (e.g.\ due to a magnetic guide
field) so that the HTD with one singled out direction can be a natural
representation. However, thorough tests would be necessary to find out
which type of decomposition is best suited for the respective physical
configurations and the differences are expected to be small.

The tree shown in Fig.~\ref{fig:ht_tree_f} can also be used
for a one- or two-dimensional velocity space by setting the
numbers of cells, ranks and values in the factor matrices $\mathrm{U}$ of one
or two dimensions to one. In a typical application, the ranks $r$ and the numbers
of cells $N$ in each velocity direction are fixed so that for
three-dimensional velocity space the
compression $C$ is given by $C = (3 N r + r^3 + r^2) / N^3$.

Low-rank Vlasov simulations can utilize such high resolutions that
the uncompressed distribution function may not fit into memory
or be too costly to compute. For this reason it is preferred to
provide the initial configuration directly in the low-rank
form instead of initializing the full distribution function and
then truncating it. Fortunately plasma simulations are often
initialized with Maxwellian distributions which can be directly
set in hierarchical Tucker format. For this purpose,
a hierarchical Tucker tensor is initialized with ranks one and
all transfer tensors and factor matrices set to one. 
Then the root tensor of $f(\mathbf{x}) \in \mathbb{H}$ (denoting the set
of hierarchical Tucker tensors by $\mathbb{H}$) is responsible for the normalization
\begin{equation}
(\mathrm{B}_{123})_{11} =  n(\mathbf{x})\,\left(\frac{m}{2 \pi T(\mathbf{x})}\right)^{d/2}
\end{equation}
with dimensionality of velocity space $d$. The factor matrices represent
the exponential distribution as
\begin{equation}
(\mathrm{U}_i)_ {k\,1} = \exp\left(-\frac{m}{2 T(\mathbf{x})}\,((v_i)_k-u_i(\mathbf{x}))^2 \right)
\quad \forall\ 1 \leq i \leq d \in \mathbb{N},\ k \in \{1,\dots,N_i\}
\end{equation}
with $(v_i)_k = v_{i,\text{begin}} + (k-1/2) \Delta v_i$.

\subsection{Computing Moments}
\label{sec:moments}

Low-rank decompositions are sometimes also called \textit{tensor networks}
because they decompose a large tensor into multiple smaller tensors
that are connected with each other. The connection can be resolved
by contractions of the multiple small tensors to again yield one
large tensor. The contraction of, for example, two third-order tensors
$\mathrm{A} \in \mathbb{C}^{\alpha_1 \times \alpha_2 \times \alpha_3}$
and $\mathrm{B} \in \mathbb{C}^{\beta_1 \times \beta_2 \times \beta_3}$
in the modes $\alpha_3$ and $\beta_1$ is denoted as $\mathrm{A} \circ_{3,1} \mathrm{B}$
and can be thought of as the inner product in these modes:
\begin{equation}
(\mathrm{A} \circ_{3,1} \mathrm{B})_{i_1,i_2,j_2,j_3}
= \sum_{s=1}^{\alpha_3} \mathrm{A}_{i_1,i_2,s} \mathrm{B}_{s,j_2,j_3}
\end{equation}
with
$\mathrm{A} \circ_{3,1} \mathrm{B} \in \mathbb{C}^{\alpha_1 \times \alpha_2 \times \beta_2 \times \beta_3}$.
This implies that modes $\alpha_3$ and $\beta_1$ must be of equal size.
Similar to Einstein notation, we will sometimes
assume contractions over identical modes when it is clear from the
context or explicitly mentioned.

Moments of the compressed distribution function can be
obtained by first contracting the factor matrices
$\mathrm{U}$ with vectors that represent the
velocity space discretization. Then the results of the
contractions are contracted with the transfer tensors $\mathrm{B}$.
This procedure can also be thought of as the implementation of the
contraction of the hierarchical Tucker tensor with a vector.

Let $\mathbf{V}_i^k \in \mathbb{R}^{N_i}$ be a vector that
represents the discretization in velocity space (to the power $k$) in direction
$i \in \{1,2,3\}$ and has elements
\begin{equation}
(V_i^k)_j = \left(v_{i,\text{begin}} + \left(j-\frac{1}{2}\right)
\Delta v_i \right)^k \Delta v_i,\quad j \in \{1,\dots,N_1\},
\label{eq:vspace_vector}\end{equation}
where $v_{i,\text{begin}}$ is the lower limit of the discrete velocity space
and $\Delta v_i$ is the cell size in direction $i$.
These vectors can be contracted with the factor matrices to yield
vectors
\begin{equation}
\mathbf{U}_i^k = \mathbf{V}_i^k \circ_{1,1} \mathrm{U}_i \in \mathbb{R}^{r_i},\quad i \in \{1,2,3\}
\end{equation}
which are then contracted with the transfer tensors.
The first moment of the distribution function is given by
\begin{equation}
n = \left( \mathbf{U}_3^{0} \circ_{1,1} (\mathbf{U}_2^{0} \circ_{1,1} \mathrm{B}_{23}) \right) \circ_{1,1}
(\mathbf{U}_1^{0} \circ_{1,1} \mathrm{B}_{123}).
\label{eq:density_ht_explicit}\end{equation}
A different notation with implicit contractions over identical modes is
\begin{equation}
n = \mathbf{U}_3^{0\ r_3}\ \mathbf{U}_2^{0\ r_2}\ \mathrm{B}_{23}^{r_2 \times r_3 \times r_{23}}
\ \mathbf{U}_1^{0\ r_1}\ \mathrm{B}_{123}^{r_1 \times r_{23} \times 1}.
\end{equation}
For better readability the tensor dimensions are omitted and the
directions are denoted by $\{x,y,z\}$ instead of $\{1,2,3\}$ so that \eqref{eq:density_ht_explicit}
is shortly written as
\begin{equation}
n = \mathbf{U}_z^0\ \mathbf{U}_y^0\ \mathrm{B}_{23}\ \mathbf{U}_x^0\ \mathrm{B}_{123}.
\end{equation}
With this notation, higher moments are given by
\begin{align}
\nonumber u_x n =& \mathbf{U}_z^0\ \mathbf{U}_y^0\ \mathrm{B}_{23}\ \mathbf{U}_x^1\ \mathrm{B}_{123}, \quad
u_y n = \mathbf{U}_z^0\ \mathbf{U}_y^1\ \mathrm{B}_{23}\ \mathbf{U}_x^0\ \mathrm{B}_{123},\\
          u_z n =& \mathbf{U}_z^1\ \mathbf{U}_y^0\ \mathrm{B}_{23}\ \mathbf{U}_x^0\ \mathrm{B}_{123},
\end{align}
and
\begin{equation}
\mathcal{P}_{xx} = \mathbf{U}_z^0\ \mathbf{U}_y^0\ \mathrm{B}_{23}\ \mathbf{U}_x^2\ \mathrm{B}_{123}, \quad
\mathcal{P}_{xy} = \mathbf{U}_z^0\ \mathbf{U}_y^1\ \mathrm{B}_{23}\ \mathbf{U}_x^1\ \mathrm{B}_{123}, \quad
          \dots
\end{equation}
and
\begin{align}
\nonumber \mathcal{Q}_{xxx} =& \mathbf{U}_z^0\ \mathbf{U}_y^0\ \mathrm{B}_{23}\ \mathbf{U}_x^3\ \mathrm{B}_{123}, \quad
          \mathcal{Q}_{xxy} = \mathbf{U}_z^0\ \mathbf{U}_y^1\ \mathrm{B}_{23}\ \mathbf{U}_x^2\ \mathrm{B}_{123}, \quad
          \dots \\
          \mathcal{Q}_{xyz} =& \mathbf{U}_z^1\ \mathbf{U}_y^1\ \mathrm{B}_{23}\ \mathbf{U}_x^1\ \mathrm{B}_{123}, \quad
          \dots
\end{align}
The computation of moments in the presented way utilizes the midpoint rule for
the integration over velocity space.

\subsection{Split-step semi-Lagrangian scheme}
\label{sec:split}

Semi-Lagrangian algorithms discretize the distribution function on a grid, solve the characterstics backward in time for the grid points and interpolate the value at the foot of the characteristics from the values at the grid points for the previous time steps. It is common to apply an operator splitting between the $\xb$ and $\vb$ advection steps. In the second-order time splitting that we use, a half velocity space
step (using $\Delta t / 2$ for the update) is followed by a full step
in position space and a second half step in velocity space. This is
also known as the leapfrog method.
 Moreover, the $\xb$ and $\vb$ advection steps are further split into three one-dimensional advection steps. The advantage of such a splitting is that the advection coefficients are then independent of the advection direction and thus the characteristics are simple.

Many of the available semi-Lagrangian interpolation methods (e.g.\ Lagrangian interpolation or spline interpolation) are applicable in the context of a low-rank Vlasov solver.
In this paper, we use the positive and flux-conservative (PFC) method from \cite{filbet-sonnendruecker-bertrand:2001} without limiters,
since this method leads to relatively few numerical oscillations. We simplify the weights from the original PFC scheme in order to make
them applicable in the low-rank solver. It is not yet clear how the slope
limiters for preserving the positivity of the distribution function
are efficiently applied when $f$ is in
a low-rank format because the cells that are arguments of the minimum function
in the limiter are not explicitly available. Therefore, the limiters are for now turned
off ($\epsilon^+ = \epsilon^- = 1$). This issue persists for all typical slope
limiters, and it is clear that it is non-trivial to prevent negative values
of the distribution function when it is not known which cells
actually violate the positivity.


Before explaining how the PFC scheme can be implemented for a low-rank representation of the form \eqref{eq:representationf}, let us briefly recall the PFC scheme. We will focus on the one-dimensional case, since we split the steps into one-dimensional interpolations. We consider the equation
\begin{equation}\label{eq:advection1d}
\partial_t f(t,x) + a \partial_x f(t,x) = 0, 
\end{equation}
for $a \in \mathbb{R}$ independent of $x$ and $t$.

In the PFC scheme the grid points are
not associated with the value of $f$ at that certain point but with the average of $f$ in
the respective cell (or the cell integral). Then the semi-Lagrangian method can be formulated
in terms of fluxes in and out of the cell.
Thus, $f_i$ is now the average value of $f$ within cell $i$ which
is given by
\begin{equation}
f_i^n = \frac{1}{\Delta x} \int^{x_{i+1/2}}_{x_{i-1/2}} f(t^{n},x)\,\D x.
\end{equation}
 The semi-Lagrangian update of the cell
integral is
\begin{equation}
\int^{x_{i+1/2}}_{x_{i-1/2}} f(t^{n+1},x)\,\D x =
\int^{X(x_{i+1/2})}_{X(x_{i-1/2})} f(t^{n},x)\,\D x
\end{equation}
where $X$ is the origin of the particle trajectory and
$x_{i-1/2} = x_{\text{begin}} + (i-1)\,\Delta x$ is the left border
of the $i$th cell with $x_{\text{begin}}$ the lower limit in $x$-direction. 
For equation \eqref{eq:advection1d}, this is given by $X(x_{i-1/2}) = x_{i-1/2} - a \Delta t$.
The flux out of
the cell is defined as
\begin{equation}
\Phi_{i+1/2}(t^n) = \int^{x_{i+1/2}}_{X(x_{i-1/2})} f(t^{n},x)\,\D x
\end{equation}
so that the semi-Lagrangian update can be written as
\begin{equation}
\int^{x_{i+1/2}}_{x_{i-1/2}} f(t^{n+1},x)\,\D x =
\Phi_{i-1/2}(t^n) + \int^{x_{i+1/2}}_{x_{i-1/2}} f(t^{n},x)\,\D x - \Phi_{i+1/2}(t^n).
\end{equation}

Let us define $\alpha_i = \frac{X(x_{i+1/2})-x_{j-1/2}}{\Delta x}$, where $j= i + \text{floor}\left( \frac{X(x_{i+1/2})-x_{i-1/2}}{\Delta x} \right)$ is the index of the cell where $X(x_{i+1/2})$ is located. Then, the interpolation weights are computed as functions of the parameter $\alpha_i$. For example for a third order interpolation stencil, the weights are then given by
\begin{align}
w^i_{-1} &= -\alpha_i\,(1-\alpha_i)\,(2-\alpha_i)\,/6 \nonumber\\
w^i_0 &= \begin{cases} (1-\alpha_i)\,\big(1 - \alpha_i\,(2-\alpha_i)\,/6 + \alpha_i\,(1+\alpha_i)\,/6 \big) & \alpha>0 \\
-\alpha_i\,\big(1 - (1-\alpha_i)\,(2-\alpha_i)\,/6 + (1-\alpha_i)\,(1+\alpha_i)\,/6 \big) & \alpha<0.
\end{cases}
 \nonumber\\
w^i_{1} &=  \alpha_i\,(1-\alpha_i)\,(1+\alpha_i)\,/6.
\end{align}
and the flux is given by
$$
\Phi_{i+1/2} = f_{j-1}^n w^i_{-1} + f_{j}^n w^i_{0} + f_{j+1}^n w^i_{1}.
$$
Note that the index $i$ of the weights can be dropped, if $\alpha$ is independent of the index $i$ which is the case when the advection coefficient (that determines $\alpha_i$) is constant along the advection direction. This is the case for our split-step Vlasov solver.

\subsection{Step in Position Space}\label{sec:stepx}

First, we discuss the $\mathbf{x}$ advection, i.e. we solve the equation
\begin{equation}
\partial_t f + \mathbf{v} \cdot \nabla_{\mathbf{x}} f = 0. 
\end{equation}
Denoting the value of $f$ at time $t^n$ by $f^n$ and the new
value at time $t^{n+1} = t^n + \Delta t$ by $f^{n+1}$ it is
\begin{equation}
f^{n+1}(\mathbf{x},\mathbf{v}) =
f^{n}\left(\mathbf{x} - \mathbf{v} \Delta t, \mathbf{v} \right).
\label{eq:semi_lagrange_update}\end{equation}
We can further split this into three one-dimensional advection problems. Let us exemplify the steps for the case of the first direction $x$. In this case, the normalized advection velocity is $ \frac{x_{i-1/2}-X(x_{i+1/2})}{\Delta x} = -\frac{\Delta t v_1}{\Delta x}$. Hence, $\alpha$ and thus the weights are depending on the first velocity dimension only. 
Let us denote by $\ib$ the three-index in position space and by $\mathbf{k}$ the three-index in velocity space. Then the flux is given by
\begin{equation}\label{eq:phi_x}
\Phi_{i_1+1/2, i_2, i_3, \mathbf{k}} = f_{j-1,i_2, i_3,\mathbf{k}}^n w_{-1}(\alpha_{k_1}) + f_{j,i_2, i_3,\mathbf{k}}^n w_{0} (\alpha_{k_1})+ f_{j+1,i_2, i_3,\mathbf{k}}^n w_{1}(\alpha_{k_1}).
\end{equation}

Note that the index $j$ depends on the $v_1$ coordinate, i.e.~on $k_1$ for the discretization. Hence, the dependence the expression \eqref{eq:phi_x} for the flux on $k_1$ is twofold: by the shift in index $i_1$ and the dependence of the weights. In order to split the dependence of the operator into a sum of Kronecker product operations, we want to construct weight vectors $W_\ell \in \mathbb{R}^{N_4}$, where $N_4$ is the number of grid points along $v_1$, such that the flux vector can be constructed as
\begin{equation}
	\Phi = \sum_{\ell} S_{\ell}^1(f) \star_4 W_{\ell}.
\end{equation} 
Here, $\star_4$ denotes the Hadamard product along direction 4. $S_{\ell}^1$ denotes the operation shifting the first index by $\ell$, i.e.
$$
\left( S_{\ell}^1(f) \right)_{\ib \mathbf{k}} = f_{\ib+(\ell,0,0), \mathbf{k}}.
$$
This operations has to be closed by suitable boundary conditions -- we use a cyclic index shifting that corresponds to periodic boundary conditions. 

 If we assume that the advection velocity is limited by one cell size, we only have to consider the two cases $j=-1,0$ for positive or negative velocities respectively. For the three-point stencil, this yields four weight vectors with the following entries for positive velocities ($\alpha$ negative)
 \begin{align}
w^{k_1}_{-2} &= 0, \nonumber\\
w^{k_1}_{-1} &= -\alpha_{k_1}\,(1-\alpha_{k_1})\,(2-\alpha_{k_1})\,/6 \nonumber\\
w^{k_1}_0 &= (1-\alpha_{k_1})\,\big(1 - \alpha_{k_1}\,(2-\alpha_{k_1})\,/6 + \alpha_{k_1}\,(1+\alpha_{k_1})\,/6 \big) \nonumber\\
w^{k_1}_{1} &=  \alpha_{k_1}\,(1-\alpha_{k_1})\,(1+\alpha_{k_1})\,/6.
\end{align}
 and the following entries for negative velocities
\begin{align}
w^{k_1}_{-2} &= -\alpha_{k_1}\,(1-\alpha_{k_1})\,(2-\alpha_{k_1})\,/6 \nonumber\\
w^{k_1}_{-1} &= \alpha_{k_1}\,\big(1 - (1-\alpha_{k_1})\,(2-\alpha_{k_1})\,/6 + (1-\alpha_{k_1})\,(1+\alpha_{k_1})\,/6 \big) \nonumber\\
w^{k_1}_{0} &=  \alpha_{k_1}\,(1-\alpha_{k_1})\,(1+\alpha_{k_1})\,/6, \nonumber\\
w^{k_1}_{1} &= 0.
\end{align}
If we want to allow for larger shifts, additional weight functions are necessary. In our implementation, we limit the shift to one cell in position space, since the time step restrictions are more severe for the step in velocity space in typical applications of the low-rank Vlasov solver.

A generalization of the method to a distribution function compressed in the form \eqref{eq:representationf} is straight-forward: The index shift is applied in $x$ direction where no compression is applied and the Hadamard product along mode 4 can be implemented by multiplying each column of the leaf matrix $U_4$ element-wise by the vector $W_\ell$.

In this case four weights can be computed once initially for all velocity space
cells, as they do not change with time. Then in every spatial step
in direction $x_j$ the neighbouring weighted hierarchical Tucker tensors
$S^1_{-2}(f) W_{-2} \dots S_{1}^1(f) W_1$ are added up with a truncation in between each
addition to compute the fluxes. Afterwards the new value of $f$ can be obtained as
\begin{equation}
f^{n+1}_i = \Phi_{i-1/2} + f^n - \Phi_{i+1/2}
\end{equation}
which again is a sum of HT tensors with a truncation after each
addition.




\subsection{Step in Velocity Space}
\label{sec:stepv}
The step in velocity space is given by the equation
\begin{equation}
\partial_t f + \frac{q}{m}\left( \mathbf{E} + \mathbf{v} \times \mathbf{B} \right) \cdot \nabla_{\mathbf{v}} f = 0.
\end{equation}
so that
\begin{equation}
f^{n+1}(\mathbf{x},\mathbf{v}) =
f^{n}\left(\mathbf{x}, \mathbf{v} -  \frac{q}{m}\left( \mathbf{E} + \mathbf{v} \times \mathbf{B} \right) \right).
\label{eq:semi_lagrange_update_v}\end{equation}
The three-dimensional advection in velocity space can also be split up into one-dimensional advections.
However, the Lorentz acceleration depends on velocity so that a symplectic second-order sequence
for a velocity space step is $\frac{1}{2}\ \text{step}\ v_x \rightarrow \frac{1}{2}\ \text{step}\ v_y \rightarrow 1\ \text{step}\ v_z
\rightarrow \frac{1}{2}\ \text{step}\ v_y \rightarrow \frac{1}{2}\ \text{step}\ v_x$
(Strang splitting). Due to the vector product, the components of the advection coefficient do
only depend on the other two velocity directions and the position, for example $a_{x} = a_{x}(\mathbf{x},v_y,v_z)$.
For a similar construction as for the position space update, the weights for the one-dimension directions are now
not only dependent one variable of the grid but on two, so the weights are matrices and not vectors anymore.
In order to avoid the additional costs related to this dependence, we further split the velocity update into
two subsequent advection steps, each depending only on one velocity space direction.
At the example of the $v_x$-step the two advection velocities are
\begin{equation}\label{eq:advection_vel_vx1}
a_{x,1} = -\frac{q}{m} (E_x/2+v_y B_z) \Delta t/\Delta v_x
\end{equation}
and
\begin{equation}
a_{x,2} = -\frac{q}{m} (E_x/2-v_z B_y) \Delta t/\Delta v_x.
\end{equation}
Splitting up the advection is possible because it is a linear operation with
no dependence on the updated velocity direction. Now the weights again
correspond to vectors and the strategy used for the step in position space
can be applied. The acceleration due to the electric field has been evenly
distributed between the two substeps which reduces time step restrictions
for large electric fields and can prevent small advection velocities in the
case of vanishing $(\mathbf{v}\times\mathbf{B})_j$.
In principle, the velocity space update can also be conducted using a single
advection step, but then the weights must be constructed in hierarchical
Tucker format and the multiplication of $f$ with the weights is an
element-wise multiplication of two hierarchical Tucker tensors. This results
in additional truncations or increased computational cost.

Let us now consider the construction of the flux for the advection by the advection velocity \eqref{eq:advection_vel_vx1}. Since we have a full grid in position space, we can consider the hierarchical Tucker tensor for each of these grid points separately
\begin{equation}\label{eq:flux_advectionv}
	\Phi_\ib = \sum_{\ell} S_{\ell}^4(f_\ib) \star_5 W_{\ell}^\ib.
\end{equation} 

Since we expect the advection velocity, normalized to the grid size, to be higher in the velocity advection step, we here restrict the shift to a maximum of two grid cells, i.e.~we construct weights for $\ell \in \{-3,-2,-1,0,1,2\}$ for the three-point stencil with three weights being zero for each coefficient. 

The Hadamard product in \eqref{eq:flux_advectionv} can be implemented as in the position step. The index shift is now along a direction that is compressed by the low-rank format. Fortunately, $f_{i-2} \in \mathbb{H}$ can
be easily obtained from $f_{i} \in \mathbb{H}$ by simply shifting the
rows of the factor matrix $\mathrm{U}$ that belongs to the respective
velocity space direction by two rows. Let $(\mathbf{U}_{k})_i$
be the $k$th row vector of the factor matrix of $f_i$ that belongs to the
updated direction and $(\mathbf{U}_{k})_{i-2}$ be the $k$th row vector
of the respective factor matrix of $f_{i-2}$, then
\begin{equation}
(\mathbf{U}_{k})_{i-2} = (\mathbf{U}_{k-2})_i \quad \forall \, k \in \{3,\dots,N_j\},
\end{equation}
where $N_j$ is the number of cells in the updated direction. The shift of rows
can be performed for all $k$ if one chooses appropriate boundary conditions.
In order to make sure that no particles are lost or created at the velocity space borders,
we implement zero-flux boundary conditions. This can be
achieved by setting in the factor matrices of the fluxes $\Phi_{i \pm 1} \in \mathbb{H}$
rows associated with the upper and lower limit of velocity space to zero
such that there is no flux out of the border cells (and also no flux from
outside of the velocity space limit).

\subsection{Dual Solver for Conservation of Particle Density}
\label{sec:dual_solver}

Every truncation of the distribution function is accompanied by a
small error which breaks the conservation properties that a Vlasov
scheme can have without low-rank approximation. In particular,
the low-rank version of the PFC scheme does no longer conserve
mass. This is of course problematic for the physical reliability
of the simulation, but also for the interplay of the Vlasov solver and the Maxwell
solver. For the solution of Maxwell's equations we employ the
finite-difference time-domain (FDTD) method. The involved
staggered grid guarantees $\nabla \cdot \mathbf{B} = 0$. However,
Gauss's law $\nabla \cdot \mathbf{E} = \rho / \epsilon_0$
will not be fulfilled when the charge density is not conserved. A simple rescaling of $f$ before the
Maxwell step in order to conserve total charge density does improve the
situation but is no sufficient solution---there are still
local errors in $\rho$.

\begin{algorithm}
\SetAlgoLined
\vspace{0.2cm}
Initial half step of the Maxwell solver\\
 \Timestep{}{
  Calculate second moment $\mathcal{P}^{n}$ from $f^n$\\
  Full Vlasov Leapfrog Step to advance to $f^{n+1}$\\
  Calculate $\mathcal{P}^{n+1}$ from $f^{n+1}$\\
  Interpolate linearly to get $\mathcal{P}^{n+1/2}$\\
  Full Runge-Kutta Fluid Step (input $\mathcal{P}$ at appropriate times)\\
  \Momentfitting{}{
    Calculate mean velocity $\mathbf{u}_{\text{V}}$ from $f$\\
    Artificial advection step with advection velocity $\mathbf{u}_{\text{V}}-\mathbf{u}_{\text{F}}$
    to enforce $\mathbf{u}_{\text{V}}=\mathbf{u}_{\text{F}}$\\
    Calculate density $n_{\text{V}}$ from $f$\\
    Multiply $f$ by $n_{\text{F}}/n_{\text{V}}$ to conserve mass\\
  }
  Full step of the Maxwell solver\\
 }
 \caption{Time stepping of the low-rank Vlasov-Maxwell solver with moment fitting}
 \label{alg:timestep_dual_lrv}
\end{algorithm}

To conserve mass in the low-rank Vlasov solver, the advection
scheme is combined with a partial differential equation (PDE) fluid solver
using moment fitting \cite{rieke-trost-grauer:2015,trost-lautenbach-grauer:2017,allmann-rahn-lautenbach-grauer:2021}.
The idea behind such a dual Vlasov solver is to transfer desirable
properties that a fluid scheme has---here the conservation of mass
and a good representation of Gauss's law---to a Vlasov scheme which
is missing those properties. Both the Vlasov solver and a fluid solver
with kinetic input operate at the same time. When the fluid equations
\eqref{eq:tenmoment_continuity} and \eqref{eq:tenmoment_movement} are
closed with help of the kinetic second moment $\mathcal{P}$ obtained from
the distribution function, they yield kinetically exact solutions
for the density $n_s$ and the mean velocity $\mathbf{u}_s$. Since the
fluid solver conserves mass and fulfills the continuity equation independent
of the conservation properties associated with $\mathcal{P}$, its kinetic solutions
for $n_s$ and $\mathbf{u}_s$ can be used to correct the respective moments
in the distribution function. This way, the resulting low-rank dual Vlasov solver
can conserve mass and take Gauss's law into account.
In the dual Vlasov solver in \cite{allmann-rahn-lautenbach-grauer:2021} the
distribution function was corrected by an exchange of the ten moment Maxwellian part of the distribution
function with the ten moment Maxwellian obtained from the fluid solver.
Thus, the information from the first three fluid equations can be utilized
so that energy is corrected and conserved.
In low-rank format this is not possible at good performance because
the ten moment Maxwellian mixes two velocity space directions
and therefore does not separate. It cannot be constructed directly
in low-rank format and a truncation starting from the full tensor
in each time step is much too costly. However, density and mean velocity can
easily be corrected. Here, the fluid solver utilizes a centrally weighted essentially
non-oscillating (CWENO) method \cite{kurganov-levy:2000} and the
third-order Runge-Kutta scheme in \cite{shu-osher:1988}.
A time step of the dual low-rank
Vlasov solver, as implemented, is described in Algorithm~\ref{alg:timestep_dual_lrv}.
First, a step of the low-rank Vlasov scheme is performed, followed by
a step of the fluid scheme with input from the distribution function.
The electric and magnetic fields obtained from the FDTD Maxwell solver
are linearly interpolated from the staggered
grid to the cell centers where they are needed by the plasma solvers.
After the plasma solver steps, mean velocity is corrected by an artificial advection step and
density is corrected by rescaling the distribution function.

As mentioned, this version of the moment fitting uses a four moment fluid solver
where the hierarchy of fluid equations is closed at the
momentum equation \eqref{eq:tenmoment_movement} using the second moment
tensor $\mathcal{P}$ obtained from the distribution function.
The density of the stand-alone low-rank Vlasov solver is not conserved
and is therefore corrected by the solution from the PDE fluid solver.
Here, we correct the mean velocity not via an exchange of Maxwellians as
in \cite{allmann-rahn-lautenbach-grauer:2021}, but through
an artificial advection step which moves the Vlasov solution towards the
solution from the PDE solver. This method was introduced in \citep{rieke-trost-grauer:2015}
for spatial coupling of Vlasov and fluid models. The artificial advection step
in velocity space is performed using the same scheme and interpolation as the regular advection
steps described in Sec.~\ref{sec:stepx} and Sec.~\ref{sec:stepv}.
For a non-positive scheme like the low-rank scheme here, the advection method has the advantage that
it does not introduce additional negativity caused by the exchange of Maxwellians.
Since the four moment fluid solver does not yield a solution for the energy
tensor, energy is not included in moment fitting and total energy is not
conserved in the resulting low-rank dual Vlasov solver. However, the low-rank
method allows much higher velocity space resolutions than the full grid solver
and therefore also has less issues with numerical heating. The design
of an energy-conserving low-rank solver will be an important aspect
of future work in this direction.

\section{Landau Damping}
\label{sec:landau_damping}

\begin{figure}
\includegraphics[width=0.9\textwidth]{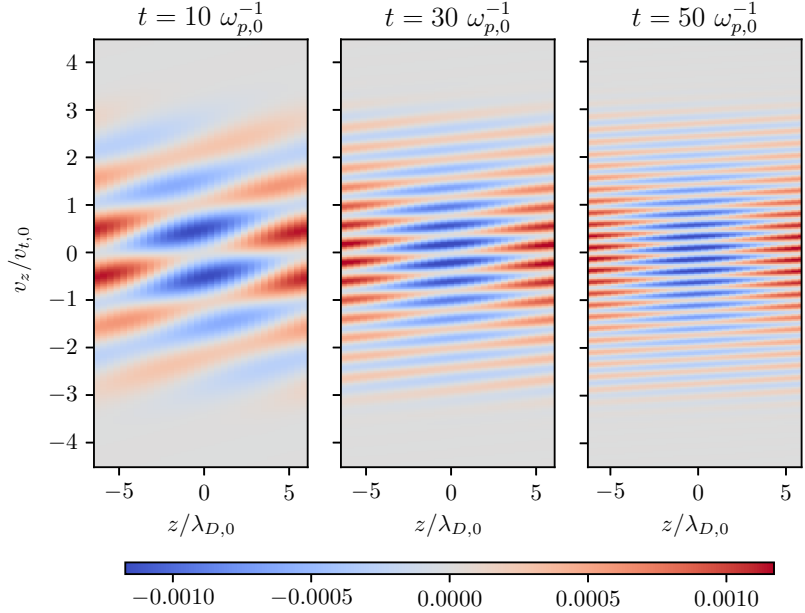}
\caption{Filamentation of the distribution function in three-dimensional Landau damping.
Shown is $f(t)-f(t=0)$ in the $z-v_z$ phase-space plane at $x=y=v_x=v_y=0$.}
\label{fig:landau_filamentation}\end{figure}

Landau damping is the predominant wave damping mechanism in
collisionless plasmas. Because it is a kinetic effect that cannot easily be
reproduced in fluid models and because an analytic solution
is known for the linear regime, it has become a standard test
case for kinetic plasma solvers. A static and spatially uniform
neutralizing ion background is
assumed and as the initial condition for the electron distribution
function we use
\begin{equation}
f_{e,0}(\mathbf{x},\mathbf{v}) = \frac{1}{(2 \pi)^{3/2}}\ \ \mathrm{e}^{\frac{-|\mathbf{v}|^2}{2}}\
    \left(1 + \alpha \sum_{j=1}^3 \cos(k_j x_j)\right)
\label{eq:initial_condition_landau}\end{equation}
which is a perturbed Maxwell distribution in three dimensions.
The setup is electrostatic and the normalization is given by: Time in units of the inverse electron plasma frequency $\omega_{p,0}^{-1}$,
length in units of the electron Debye length $\lambda_{D,0}$, velocity in units of the electron thermal
velocity $v_{t,0}$, mass in units of the electron mass $m_e$, temperature in units of the initial electron temperature $T_0$,
charge in units of the ion charge $q_i$ and vacuum permittivity $\epsilon_0 = 1$.
Here, it is $k_j = 0.5\,\lambda_{D,0}^{-1}$
for all $j$ and the perturbation is set to $\alpha = 0.01$.
The physical domain goes from $x_j=-2\pi\,\lambda_{D,0}$ to $x_j=2\pi\,\lambda_{D,0}$
and the velocity space from $v_j=-4.5\,v_{t,0}$ to $v_j=4.5\,v_{t,0}$ in each direction $j$.
Speed of light is set to $10\,v_{t,0}$. The fields are evolved by the Maxwell solver
(despite the electrostatic nature of the problem) in order to demonstrate that the
presented six-dimensional Vlasov--Maxwell scheme can capture electrostatic Landau damping accurately.

\begin{figure}
\centering \includegraphics[width=0.75\textwidth]{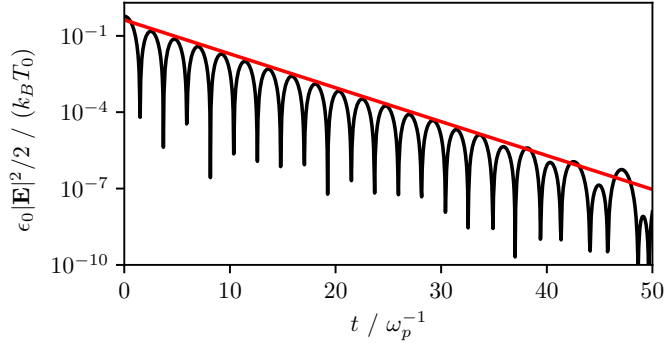}
\caption{Damping of the electric wave energy in three-dimensional Landau damping.
Red: Analytic damping rate.}
\label{fig:landau_damping}\end{figure}

The three-dimensional initialization that we chose separates in velocity
space and is therefore perfectly suited for the low-rank solver. We perform
a simulation with a resolution of $32^3$ in position space and
$256^3$ in velocity space. Due to the separability it is sufficient to
use rank one so that there is a lossless compression of the distribution function
to $1/20\,000$ of the full-grid data. The simulation can be run on a single
modern CPU whereas the full-grid version would require a large supercomputer.
Admittedly, this is the most ideal situation for the low-rank solver, but
it is nevertheless worthwhile to demonstrate that the high velocity
space resolution can be utilized despite the massively reduced
degrees of freedom. This is shown in Fig.~\ref{fig:landau_filamentation}
where the deviation of the distribution function from the initial
value is given for different times. Due to the different speeds of
the particles in the Maxwell distribution,
increasingly small scale structures develop which is known as the
filamentation of the distribution function. The filamentation
is accurately resolved by the high velocity space resolution despite the
compression. In Fig.~\ref{fig:landau_damping} the damping of the
electric energy is shown next to the analytic damping rate of
$\gamma = 0.1533\,\omega_{p,0}$. The simulation matches this
rate, only at later times the discretization
error in position space begins to dominate the small remaining
perturbation which is expected at a resolution of $32^3$. Within the simulated time span there is no recurrence effect
which would occur if the filamentation became finer than the velocity resolution.

\section{Orszag-Tang Turbulence}
\label{sec:orszag_tang}

In this section we compare the low-rank Vlasov solver to the full grid version in
a well-known plasma turbulence setup, the Orszag-Tang vortex. We choose parameters
as in \cite{groselj-cerri-navarro:2017} apart from a lower ion-electron mass ratio (here ${m_i/m_e = 25}$).
The initial magnetic field is ${B_y = -\delta_B \sin(2 \pi z / L)}$, ${B_z = \delta_B \sin(4 \pi y / L)}$
and ${B_x = 1 B_0}$ and the velocities are ${u_{y,s} = -\delta_u \sin(2 \pi z / L)}$,
${u_{z,s} = \delta_u \sin(2 \pi y / L)}$, ${u_{x,i} = 0}$ and
${u_{x,e} = -\frac{2 \pi}{L} \delta_B \mu_0 (2 \cos(4 \pi y / L) + \cos(2 \pi z / L))}$.
The perturbation has a magnitude of ${\delta_u = 0.2\,v_{A,0}}$ and ${\delta_B = 0.2\,B_{0}}$.
The current in $x$-direction is carried by the electrons. The electric field is
${E_y = -\delta_u B_0 \sin(2 \pi y / L)}$, ${E_z = -\delta_u B_0 \sin(2 \pi z / L)}$ and ${E_x = 0}$.
The initial density is uniform ${n_s = n_0}$ apart from a small perturbation
added to same parts to the electron and ion densities in order to satisfy Gauss's law.
The temperature ratio is ${T_i/T_e = 1}$
with ${\beta_i = 2 \mu_0 n_0 k_B T_i / B_0^2 = 0.1}$ and the speed of light is ${c = 18.174\,v_{A,0}}$.
The distribution function is initialized as a Maxwellian distribution based on the aforementioned
quantities. The spatial resolution is chosen as $128^2$ in order to resolve electron inertial length
in the periodic domain of size ${L_y = L_z = L = 8 \pi\,d_{i,0}}$.
The velocity space extends from $-9\,v_{A,0}$ to $9\,v_{A,0}$ for the electrons and
from $-2\,v_{A,0}$ to $2\,v_{A,0}$ for the ions.

In electromagnetic setups like Orszag-Tang turbulence (and also GEM reconnection
in the next section) we choose the following normalization:
Length in units of the ion inertial length ${d_{i,0}}$ based on
density ${n_0}$, velocity in units of the ion Alfv\'{e}n velocity ${v_{A,0}}$ based
on the magnetic field ${B_0}$, time in units of the inverse of the ion
cyclotron frequency ${\Omega_{i,0}^{-1}}$, mass in units of the ion mass ${m_i}$,
electric charge in units of the ion charge ${q_i}$,
vacuum permeability ${\mu_0 = 1}$ and Boltzmann constant ${k_B = 1}$.

\begin{figure}
\includegraphics[width=\textwidth]{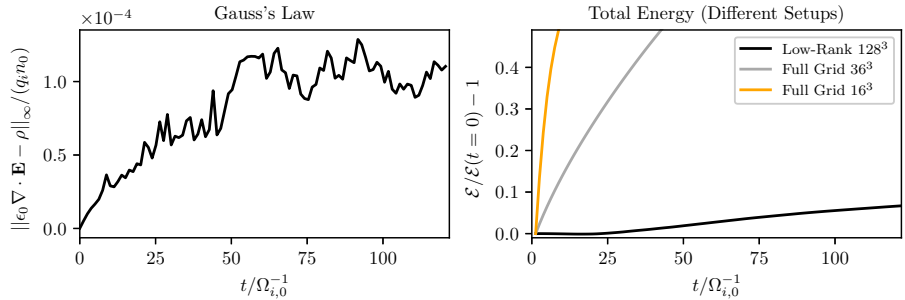}
\caption{Evolution of the error in Gauss's law and total
energy conservation over time in the Orszag-Tang
turbulence simulation.}
\label{fig:orszag_tang_mass_gauss_energy}\end{figure}

The low-rank simulation uses a velocity space resolution of $128^3$ cells and
rank $r=8$. This results in a compression of 1:574, i.e.\ a compression
to less than 0.18\% of the full data. Compared to a full grid simulation that
is resolved by $128^3$ cells, the low-rank simulation achieves a speed-up by
a factor of 70. Here, we compare the low-rank simulation (i) to a full grid simulation that takes
the same computational time (wall time) per step and (ii) to a full grid simulation that uses
the same degrees of freedom (DOF). The respective velocity space resolutions are
$36^3$ (same wall time) and $16^3$ (same DOF). Note that the implementation of the
hierarchical Tucker algorithms used by the low-rank solver still has lots of room
for performance improvement, for example by avoiding unnecessary copy operations
and allocations/deallocations. The full grid solver on the other hand is highly optimized already.
For an otherwise fair comparison, we utilize in the full-grid simulations the advantages that an explicitly
available grid brings: The positivity preserving limiters are turned on and
the velocity splitting is realized via the backsubstitution method \cite{schmitz-grauer:2006-2}.
The full grid simulations are performed with the \textit{muphy2} multiphysics plasma simulation code
\cite{allmann-rahn-lautenbach-grauer:2021,allmann-rahn-lautenbach-grauer-etal:2021}.

\begin{figure}
\includegraphics[width=\textwidth]{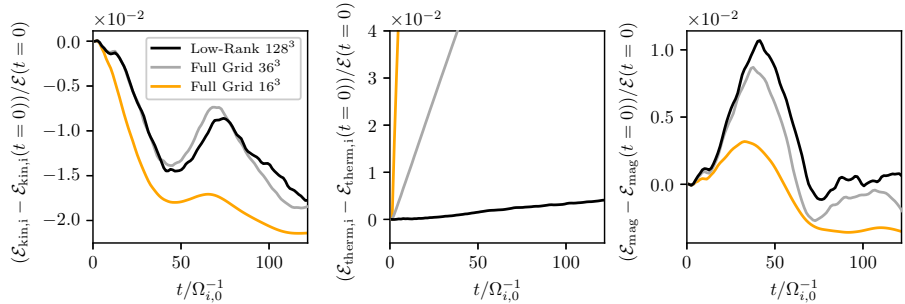}
\caption{Change of energy partition between ion kinetic energy, ion
thermal energy and magnetic energy in the low-rank Vlasov solver
compared to the full grid version at the same wall time ($36^3$)
and the same degrees of freedom ($16^3$).}
\label{fig:orszag_tang_energy_partition}\end{figure}

First, we will discuss some features of the low-rank solver.
The low-rank approximation leads to non-conservation of the distribution function and in
consequence non-conservation of total mass. This can be corrected by means of moment fitting as presented in
Sec.~\ref{sec:dual_solver}. The mass correction on basis of a fluid solver with kinetic input enables the low-rank Vlasov-Maxwell solver to
self-consistently take Gauss's law into account, without the need for divergence cleaning
which was used for example in \cite{einkemmer-ostermann-piazzola:2020}.
The FDTD Maxwell solver evolves the fields according to Amp\`{e}re's law and Faraday's law. These two equations are
sufficient if $\nabla \cdot \mathbf{B}=0$ and Gauss's law $\nabla \cdot \mathbf{E}=\rho/\epsilon_0$
are fulfilled initially and the  plasma solver satisfies some requirements: It needs to provide the appropriate
current density as an input to Amp\`{e}re's law on the staggered grid and has to
fulfill the continuity equation for charge.
The current density on the staggered grid can be provided by the reconstruction that is
performed during the solution of the fluid equations, and the continuity equation is fulfilled
in form of the first fluid equation.
In Fig.~\ref{fig:orszag_tang_mass_gauss_energy} the errors in
Gauss's law (left) and in energy conservation (right) are shown.
As becomes apparent, there is still a small error present in Gauss's law which is related to the different
time-stepping in the fluid solver and the Maxwell solver. In order to further improve the representation of Gauss's law,
the FDTD Maxwell solver could for example be replaced by a solver that uses Runge-Kutta
time-stepping like the fluid solver. However, the FDTD method has several desirable properties
(it is energy-conserving, symplectic, robust and stable and has low numerical dissipation)
and for many applications the obtained accuracy in Gauss's law is sufficient.

As discussed, the current implementation of the low-rank solver, as well as the positivity
preserving full grid scheme it is based on, are not energy-conserving. Analytically, total
energy ${\mathcal{E} = \int\,\sum_s (m_s n_s |\mathbf{u}_s|^2/2 +  n_s k_B T_s) +
|\mathbf{B}|^2/(2 \mu_0) + \epsilon_0 |\mathbf{E}|^2/2}\,\D \mathbf{x}$
is conserved in the Vlasov-Maxwell system of equations (in for example an infinite or
periodic domain). The high velocity
space resolutions that are possible with the low-rank solver lead to lower errors
in energy conservation as shown in Fig.~\ref{fig:orszag_tang_mass_gauss_energy}.
There, the low-rank solver with resolution $128^3$, $r=8$ is compared to the full grid solver
at resolutions of $36^3$ (same computational time) and $16^3$ (same degrees of freedom).
The significantly reduced numerical heating in the low-rank solver is a consequence
of improved numerical accuracy in addition to the improved physical accuracy that the
higher resolution provides. Nevertheless,
a more exact energy conservation is desirable and improvements to the low-rank solver
in this direction are necessary in the future.

\begin{figure}
\includegraphics[width=\textwidth]{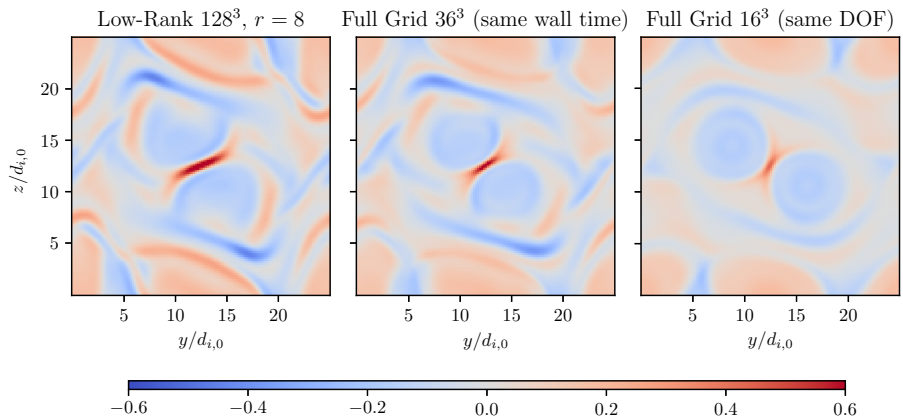}
\caption{Out-of-plane current density at $t=62.83\,\Omega_{i,0}$ in the
low-rank simulation and comparable full grid simulations.}
\label{fig:orszag_tang_mass_jx}\end{figure}

The comparison between the three configurations is continued in
Fig.~\ref{fig:orszag_tang_energy_partition} where the evolution of the energy partition
is shown, in particular the change of ion kinetic energy
$\mathcal{E}_{\text{kin},i} = \int m_i n_i |\mathbf{u}_i|^2/2\,\D \mathbf{x}$, ion thermal energy
$\mathcal{E}_{\text{therm},i} = \int n_i k_B T_i\,\D \mathbf{x}$ and magnetic energy
$\mathcal{E}_{\text{mag}} = \int |\mathbf{B}|^2/(2 \mu_0)\,\D \mathbf{x}$ over time.
The results from the highly-resolved low-rank simulation are in good agreement with published
results from fully kinetic simulations \cite{groselj-cerri-navarro:2017,allmann-rahn-lautenbach-grauer:2021}
whereas in the comparable full grid simulations the numerical heating of the PFC Vlasov scheme in combination
with low resolutions contaminates the results. The same holds for the spatial structure
of the turbulent fields shown at the example of the out-of-plane current density $j_x$ in
Fig.~\ref{fig:orszag_tang_mass_jx}. Due to numerical dissipation, the full grid simulations miss details that the
low-rank simulation captures. There is good agreement of the current density's spatial structure
with the aforementioned published PIC and Vlasov results in the case of the low-rank solver.
This indicates that despite the exceptionally high compression no substantial kinetic information is lost.

\section{GEM Reconnection}
\label{sec:gem}

A standard test setup for magnetic reconnection is given by the Geospace
Environmental Modeling (GEM) challenge \cite{birn-drake-shay-etal:2001}.
There, the initial configuration is a Harris sheet with density
${n_s = n_{0} \sech^{2}(z/\lambda) + n_{b}}$ and magnetic field
${B_{y} = \tanh(z/\lambda) B_0 + \delta B_{y}}$, ${B_{z} = \delta B_{z}}$.
The background density is ${n_{b} = 0.2\,n_0}$ and the half-width of the
current sheet is ${\lambda = 0.5\,d_{i,0}}$. The uniform temperature is defined according to
${n_{0} k_{B} (T_{e}+T_{i}) = B_{0}^{2} / (2 \mu_{0})}$, ${T_{i}/T_{e} = 5}$.
The reconnection process is initiated by a perturbation
${\delta B_{y} = -\psi_0 \pi/L_z \cos(2 \pi y / L_y) \sin(\pi z / L_z)}$,
${\delta B_{z} = \psi_0 2\pi/L_y \sin(2 \pi y / L_y) \cos(\pi z / L_z)}$ with ${\psi_0 = 0.1 B_0 d_{i,0}}$.
The magnetic field gradients are associated with a current density which
is distributed among electrons and ions according
to ${u_{x,i}/u_{x,e} = T_i/T_e}$. The ion to electron mass ratio is ${m_i / m_e = 25}$ and
the speed of light is ${c = 20\,v_{A,0}}$.
A Maxwellian distribution computed from aforementioned quantities serves as the initial
condition for the particle distribution function.
We resolve the domain ${L_{y} \times L_{z} = 8\pi d_{i,0} \times 4\pi d_{i,0}}$
by ${256 \times 128}$ cells. The boundary conditions are periodic in $y$-direction,
at the $z$-boundaries there are reflecting walls for particles that are conducting for fields
and in $x$-direction there is no spatial variation.
We simulate only the lower right quarter of the domain and utilize symmetries of the setup
in order to save computational time. Electron velocity space ranges from
${-10\,v_{A,0}}$ to ${10\,v_{A,0}}$ and ion velocity space from
${-5\,v_{A,0}}$ to ${5\,v_{A,0}}$.

\begin{figure}
\includegraphics[width=\textwidth]{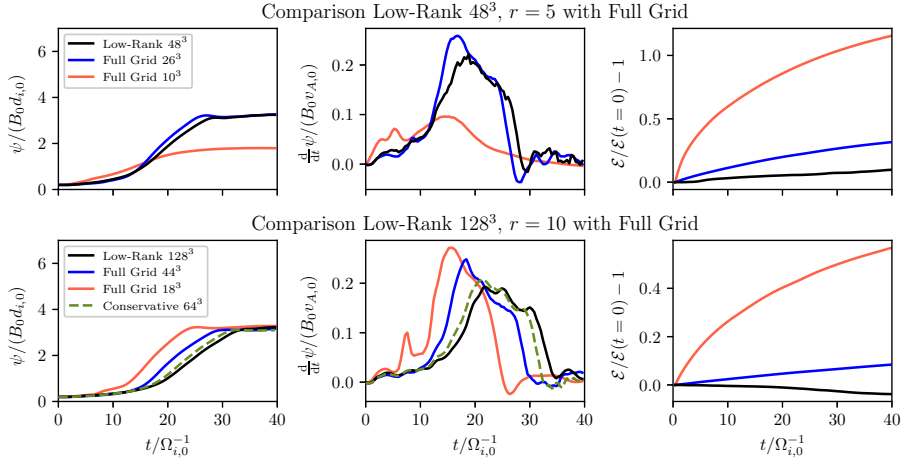}
\caption{Comparison of solver configurations: Reconnected flux $\Psi$, reconnection rate $\frac{\D}{\D t} \Psi$, and
the error in energy conservation over time in simulations of magnetic reconnection.}
\label{fig:gem_recorate_energy}\end{figure}

In this reconnection setup we test the low-rank Vlasov solver with different
ranks against comparable full grid solutions with the same wall time and
degrees of freedom, respectively, as well as a well-resolved full grid Vlasov
simulation that uses an energy conserving scheme.
The first low-rank configuration uses $48^3$ cells in velocity space and
a rank $r=5$ which corresponds to a compression of 1:127 (0.78\% of the full data).
It is compared to full grid simulations with a resolution of $10^3$
for the same degrees of freedom and $26^3$ for the same wall time. The
second low-rank configuration is resolved by $128^3$ cells with rank
$r=10$. Then the compression is 1:424 (0.24\% of the full data) so
that we achieve a speed-up by a factor of 40 compared to a full grid
simulation with $128^3$ cells. Here, the same degrees
of freedom and same wall time counterparts on the full grid use $18^3$ cells
and $44^3$ cells.
Finally, we also performed a full grid simulation that uses the positivity
preserving and energy conserving dual Vlasov solver
from \cite{allmann-rahn-lautenbach-grauer:2021} and a resolution of $64^3$ cells.

\begin{figure}
\includegraphics[width=0.9\textwidth]{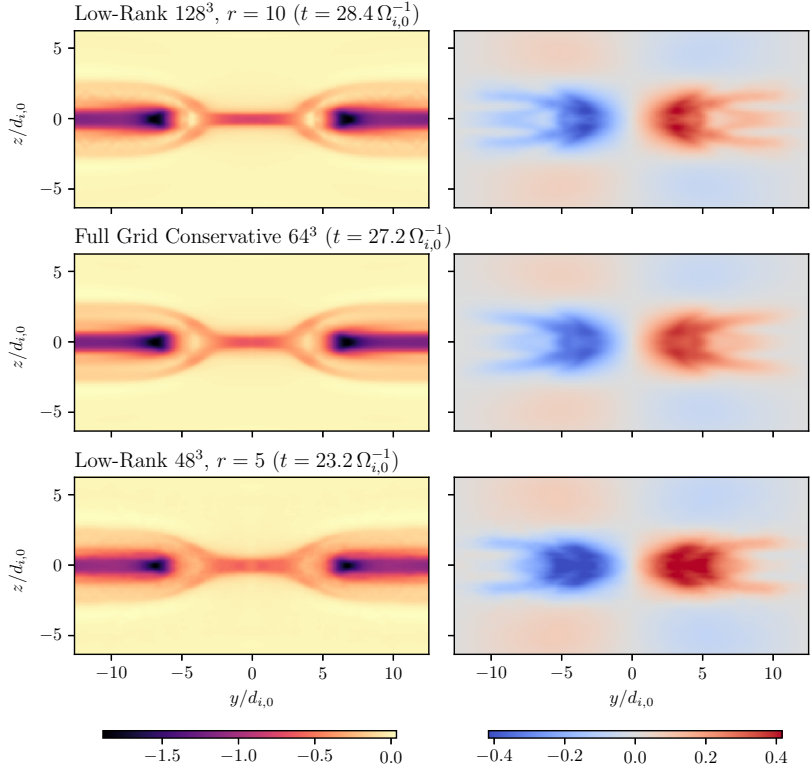}
\caption{Low-rank and full grid solutions for the out-of-plane current
density $j_x/(n_0 v_{A,0})$ (left) and ion outflow velocity $u_{y,i}/v_{A,0}$ (right) compared between different ranks
and the full grid.}
\label{fig:gem_jx_uyi}\end{figure}

\begin{figure}
\includegraphics[width=0.9\textwidth]{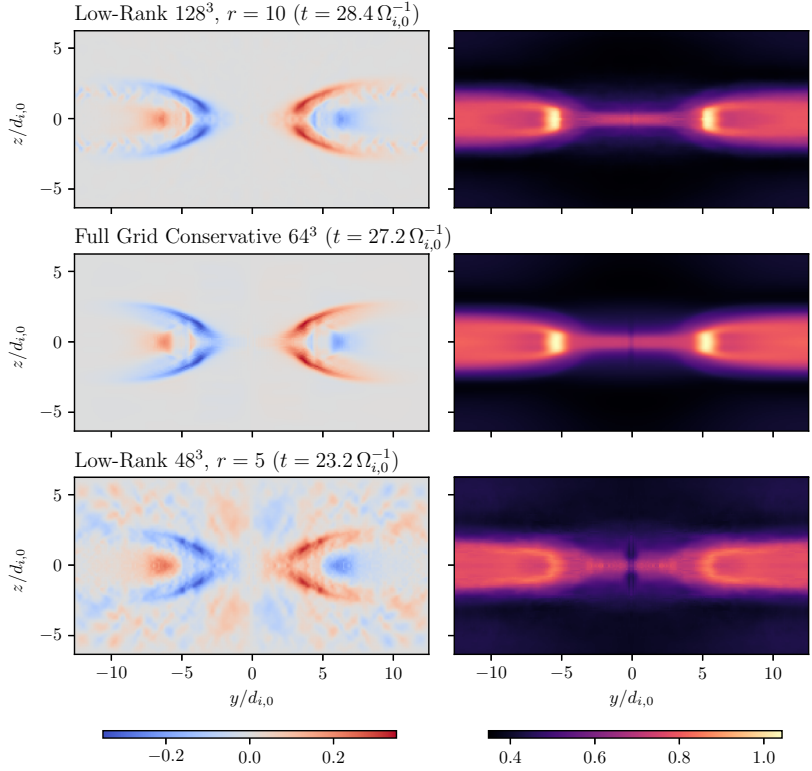}
\caption{Low-rank and full grid solutions for the hall electric
field $E_y/(B_0 v_{A,0})$ (left) and the ion temperature $T_i/(m_i v_{A,0}^2 / k_B)$ (right).}
\label{fig:gem_Ey_Ti}\end{figure}

\begin{figure}
\includegraphics[width=0.9\textwidth]{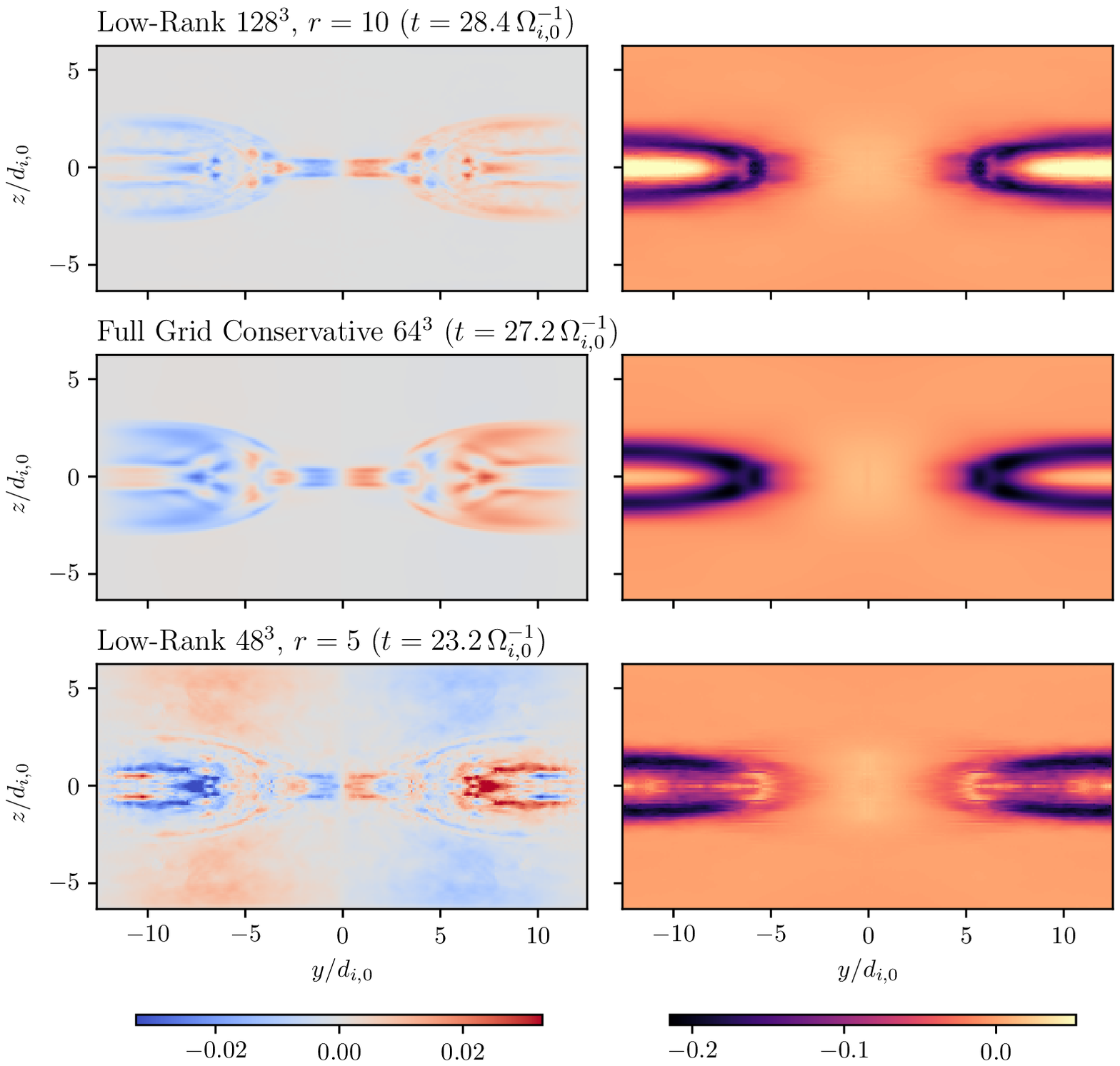}
\caption{Low-rank and full grid solutions for the electron heat flux
component $Q_{xxy,e}/(m_i n_0 v_{A,0}^3)$ and the ion heat flux component
$Q_{xyy,i}/(m_i n_0 v_{A,0}^3)$.}
\label{fig:gem_Qxxye_Qxyyi}\end{figure}

These different configurations are compared in Fig.~\ref{fig:gem_recorate_energy}
concerning the evolution of reconnected magnetic flux $\Psi$ and reconnection rate
$\frac{\D}{\D t} \Psi$ next to the error in energy conservation. The reconnected
flux is here the integral over $B_z$ at $z=0$ from the center to the right border of the domain
$\Psi = \int_0^{4 \pi} B_z(y,z=0)\,\D y$. The upper row in Fig.~\ref{fig:gem_recorate_energy}
shows the $r=5$ case and the corresponding full grid runs. At the same degrees
of freedom as the low-rank simulation, the full grid solver fails to appropriately
represent the reconnection process. The better velocity space resolution furthermore
results in a lower peak reconnection rate. In the lower row the results from the
$r=10$ simulation are shown. Again there is the tendency that the peak reconnection
rate decreases at better velocity space resolutions. The qualitative
development is similar between the $128^3$ low-rank simulation and the $64^3$
energy conserving full grid simulation. There is some numerical
cooling in the low-rank simulation that leads to a later onset of
reconnection. The slightly lower peak reconnection rate in the low-rank
case likely results from the better resolution.

The effect of compression (low-rank approximation) can best be seen in a
direct comparison of physical quantities between simulations of different
rank and a full-grid simulation. For this purpose, various quantities
are shown in the Figures \ref{fig:gem_jx_uyi}--\ref{fig:gem_Qxxye_Qxyyi}
for the $r=10$ low-rank simulation ($128^3$), the energy conserving
full grid simulation ($64^3$) and the $r=5$ low-rank simulation ($48^3$). The snapshots
are taken at times where $\Psi = 2.5 B_0 d_{i,0}$ so that the reconnection
is in a similar state. The reconnection current sheet density $j_x$ in
Fig.~\ref{fig:gem_jx_uyi} typically gives a good overview over the reconnection as
it visualizes both the X- and the O-line. There is excellent agreement between
the three configurations which is especially encouraging in the $r=5$ case
which uses very few degrees of freedom (870 DOF for the velocity space). On the
right hand side of Fig.~\ref{fig:gem_jx_uyi} the ion outflow velocity $u_{y,i}$ is
shown. It is mostly identical in the $r=10$ and full grid cases, but somewhat
overestimated by the $r=5$ run. The low-rank $r=10$ simulation yields
a slightly sharper and more detailed picture than the full grid simulation.

The Hall reconnection field $E_y$ is given in Fig.~\ref{fig:gem_Ey_Ti} next
to the ion temperature $T_i$. In both quantities, but especially in the
electric field, two characteristics of the low-rank solver are visualized in
the $r=5$ case. First, a low rank is associated with noise caused by the
round-off error during the truncation of SVDs. Second, because the solver
does not preserve positivity of the distribution function, numerical oscillations
may be present. However, both effects are relatively small in this case.
In the $r=10$ case there is almost no noise, but naturally some numerical
oscillations are visible in the electric field. Ion temperature matches
for the $r=10$ and the full grid simulation and its structure is also
caught by the $r=5$ simulation.

Finally, two heat flux components---one for the electrons and one for
the ions---are shown in Fig.~\ref{fig:gem_Qxxye_Qxyyi}. The components
of the heat flux are very sensitive to the representation of the velocity
distribution and are a good indicator of how much information is lost due
to the compression of the distribution function. It can be seen that
the detailed structure in the $Q_{xxy,e}$ component agrees accurately
between the $r=10$ low-rank solution and the full grid solution. Even
the $r=5$ solution captures the heat flux very well although the magnitude is
overestimated in some regions and the heat flux is overall somewhat
noisy. Nevertheless, this shows a very important and encouraging
characteristic of the low-rank solutions: At low ranks they yield a
less accurate picture of the correct physics rather than an
accurate picture of incorrect (non-kinetic) physics as for example
fluid approximations do. At high ranks on the other hand, almost
no kinetic information is lost. Similar conclusions can also be
drawn from the shown ion heat flux component $Q_{xyy,i}$. Again,
there is good agreement between the $r=10$ low-rank simulation
and the conservative full grid simulation and a decent representation
in the $r=5$ case.

\section{Conclusions}
\label{sec:conclusions}

A low-rank solver for the full six-dimensional Vlasov-Maxwell equations
was presented and tested in benchmark problems of plasma turbulence
and magnetic reconnection. Massive compression of the
distribution function to for example $1/500$ of the
full data is possible without significant loss of information.
With the current version of the code speed-ups of up to
70 have been achieved in realistic setups. Compared to full grid
simulations using the same numerical scheme and the same
computational time or degrees of freedom, the low-rank
simulations yield much more accurate results.
The dual Vlasov (moment fitting) method---where the Vlasov solver is complemented
by a fluid solver with kinetic moment input---makes the low-rank
solver mass conservative and enables a good representation
of Gauss's law.

Generally, the development of low-rank Vlasov solvers
is still a rather new field of research with enormous potential.
One of the most important extensions of the low-rank solver
would be the introduction of limiters to preserve the
positivity of the compressed distribution function. Equally
important is a good conservation of total energy and momentum.

The implementation of the low-rank solver is fully parallelized
using MPI and a domain decomposition approach and scales
well on large HPC CPU clusters. It would be desirable to be able to utilize
GPUs which is possible but highly non-trivial \cite{boukaram-turkiyyah-keyes:2019}.
The currently used implementation of the hierarchical Tucker operations serves more as a prototype implementation and
leaves room for optimizations and performance improvements since
no special care has been taken so far to avoid data copying and
allocations/deallocations.

In the future, the rank of the approximation may depend on position space
so that high ranks are only used where they are required to capture
physical effects. With efficient load balancing such an approach
can lead to significant performance gains in large-scale simulations.

\section*{Acknowledgments}

We gratefully acknowledge the Gauss Centre for Supercomputing e.V. (www.gauss-centre.eu) for funding this project by providing computing time
through the John von Neumann Institute for Computing (NIC) on the GCS Supercomputer JUWELS at J\"ulich Supercomputing Centre (JSC). Computations were
conducted on JUWELS-booster [J\"ulich Supercomputing Centre, 2019] and on the DaVinci cluster at TP1 Plasma Research Department. F.A. was supported
by the Helmholtz Association (VH-NG-1239).

\bibliography{bibliography}

\end{document}